\documentclass[12pt]{article}
\usepackage[a4paper,margin=1in]{geometry}
\usepackage{amsmath, amssymb}
\usepackage{graphicx}
\usepackage{natbib}
\usepackage{hyperref}
\usepackage{titlesec}
\usepackage{setspace}
\titleformat{\section}[block]{\Large\bfseries}{\thesection.}{0.5em}{}
\titleformat{\subsection}[block]{\large\bfseries}{\thesubsection.}{0.5em}{}

\title{Rational Miner Behaviour, Protocol Stability, and Time Preference:\\ An Austrian and Game-Theoretic Analysis of Bitcoin's Incentive Environment}
\author{Craig S. Wright\\
University of Exeter\\
\texttt{cw881@exeter.ac.uk}}
\date{}

\begin{document}
\maketitle

\begin{abstract}
\noindent This paper integrates Austrian capital theory and game theory to analyse the economic implications of protocol mutability in blockchain systems. Focusing on the strategic behaviour of miners, we demonstrate how institutional instability elevates time preference, undermines calculability, and transforms entrepreneurial action into opportunistic rent-seeking. Through a formal game-theoretic lens, we show that mutable rules destabilise cooperative equilibria and incentivise meta-strategic manipulation. By contrast, protocol immutability functions as a synthetic constitutional constraint—anchoring expectations, enabling rational investment, and sustaining the economic time structure necessary for capital-intensive production. Using the empirical divergence between the original Bitcoin protocol and BTC Core as a case study, we reveal how discretionary governance erodes institutional order. We argue that restoring protocol fixity is essential not only for economic coherence but for the viability of blockchain as an investment-grade infrastructure. This is not a matter of technical preference but of foundational economic logic.
\end{abstract}

\begin{center}
\end{center}

\noindent \textbf{Keywords:} Austrian economics; game theory; time preference; blockchain governance; institutional stability; Bitcoin; protocol immutability; rent-seeking; calculability; capital investment

\section{Introduction}

The economic architecture of Bitcoin offers a natural experiment in rule-based institutional design, where protocol immutability substitutes for legal enforcement and miner incentives emerge from algorithmic constraints rather than legislative fiat. Yet while neoclassical models treat miner strategy as a static optimisation over immediate rewards, such approaches neglect the role of institutional calculability in shaping temporal decision-making. This paper introduces a hybrid framework—melding Austrian time preference theory with dynamic game theory—to examine how shifts in protocol credibility recalibrate strategic horizons. By formalising the link between institutional mutability and discounting behaviour, we demonstrate that rational miners facing discretionary rule change internalise short-termism, undermining capital coordination and diminishing systemic value. In this way, Bitcoin’s protocol is not merely a technical standard, but a constitution whose credibility defines the scope of long-term strategic cooperation.

\subsection{Motivation and Scope}

The economic design of Bitcoin and similar systems remains incompletely theorised within a formal Austrian framework, particularly with respect to intertemporal strategic behaviour and institutional evolution. This paper aims to fill this lacuna by integrating Austrian capital theory—rooted in Böhm-Bawerk’s notion of roundaboutness and Misesian time preference—with the rigorous formalism of non-cooperative game theory, in order to construct a unified account of rational miner behaviour under divergent institutional regimes: fixed-rule protocols versus malleable, developer-controlled systems.

The motivation arises from an observed contradiction. The protocol, as defined in the original Bitcoin whitepaper, operates analogously to a Hayekian spontaneous order: a stable institutional framework upon which economic actors—here, miners—project long-term strategies. Any alteration to the protocol introduces institutional uncertainty akin to regime instability in fiat systems, disrupting capital structure alignment and encouraging shorter production stages. This incentivises short-term rent extraction over long-term value maximisation, not due to individual irrationality, but as an emergent property of strategic interaction under unstable rule regimes. It is in this sense that protocol mutability becomes equivalent to monetary manipulation, distorting temporal coordination among capital-allocating actors and yielding behaviour that would be deemed suboptimal in a regime of predictable, stable rules.

Through formal game-theoretic models incorporating infinite-horizon repeated games and subgame perfect equilibrium, this paper demonstrates that the Nash equilibria under fixed-rule protocols align with the Austrian ideal of deferred gratification, high time preference stability, and capital accumulation. In contrast, when the rules are subject to discretionary changes by development cartels or external governance actors, the resultant game space includes equilibria marked by higher strategy discounting, speculative allocation, and increased incentive to defect from cooperative norms (e.g. honest block creation). We show that such environments mirror time-inconsistent policy regimes and yield dynamically inconsistent equilibria, as forecasted in Kydland and Prescott's theory of rules versus discretion.

This paper’s scope, then, is twofold: firstly, to define and model the economic logic by which rational actors, situated in a blockchain protocol environment, adjust their strategies according to the credibility and immutability of institutional rules; and secondly, to derive implications for the design of future blockchain protocols and governance systems. We assert that only by embedding credible commitment devices—immutability of protocol, enforceability of economic contracts, and exogenous constraints on developer discretion—can one create a game structure where long-run investment in network infrastructure, transaction processing, and fee-based revenue supersede extractive, speculative mining strategies.

This project situates itself at the intersection of institutional economics, computational contract theory, and Austrian capital theory. It responds to and extends the work of scholars such as Selgin, White, and Salerno, by introducing formal mechanisms to the Misesian insight that economic coordination requires not only the ability to plan, but the stability of institutional horizons within which plans are enacted. In this context, Bitcoin’s protocol is not merely code but a constitution—a binding commitment that transforms the strategic landscape from a reactive series of short games into an ongoing, rationally forecastable process of productive coordination.

\subsection{Literature Context}

The Austrian School of Economics has long maintained that calculability within a stable institutional framework is a prerequisite for rational economic planning. Mises \cite{mises1949human} contends that time preference is a universal category of human action—manifest in all economic decisions—and that the temporal structure of production necessitates predictable rules for intertemporal coordination. Böhm-Bawerk’s theory of capital \cite{bohm1891capital} extends this insight by demonstrating that longer, more productive “roundabout” methods of production emerge only when individuals exhibit lower time preference and trust in the continuity of the institutional environment. Rothbard \cite{rothbard2004man} reinforces this principle, emphasising that any distortion in time preference—be it through monetary manipulation or institutional volatility—undermines capital accumulation and leads to malinvestment.

In contrast, neoclassical treatments of miner incentives, particularly in the context of Proof-of-Work (PoW) systems, typically model behaviour within static or Markovian environments. Miner strategies are often represented via stage games or repeated games wherein players maximise expected utility subject to instantaneous reward structures. Models grounded in Nash equilibrium, such as those derived from Osborne and Rubinstein’s framework \cite{osborne1994course}, formalise the incentives of rational miners as functions of computational expenditure, expected reward, and block propagation probabilities, largely abstracted from institutional change. The dominant assumption is one of structural stationarity: the protocol’s rules are fixed or exogenously stable.

However, this assumption collapses under real-world dynamics. Protocol mutability—introduced via soft forks, hard forks, or discretionary developer governance—renders the rule environment non-stationary. Yet extant game-theoretic treatments omit the endogenous impact of such rule variability on time preference and strategic horizon formation. There exists a lacuna in the literature: a time-preference-aware formal model that internalises institutional mutability as an endogenous variable shaping player strategy. The Austrian insight that calculability is a function of institutional predictability remains unformalised in computational economics. Meanwhile, the prevailing models fail to account for dynamic shifts in miner expectations when the rules governing reward structures are open to discretionary change.

This paper aims to bridge this gap by constructing a game-theoretic model in which time preference is treated not as an exogenous constant, but as a function of protocol credibility. In so doing, it synthesises Austrian capital theory with strategic formalism, providing a tractable framework for evaluating how miners recalibrate intertemporal strategies in response to mutable versus immutable rule regimes. The result is a novel contribution that situates Bitcoin protocol governance within the broader literature of institutional economics, rational expectations, and dynamic game theory.

\subsection{Cryptoeconomic Order and the Necessity of Fixed Rules}

This paper contends that the economic viability of blockchain systems rests not merely on cryptographic primitives or distributed architecture, but on the institutional permanence of inviolable rules. When protocol mutability is permitted, even nominally decentralised systems are susceptible to capture by coordinated minorities. Small, socially influential developer groups or corporate actors begin to exercise de facto control over the protocol agenda, enshrining complexity that obfuscates rule clarity and reinforces their central role. This process mirrors institutional centralisation in monetary systems, where discretionary control enables rent-seeking and bureaucratic layering. As protocol complexity grows, so too does the asymmetry of information and access, displacing spontaneous order with curated discretion. In contrast, a cryptoeconomic system governed by fixed, algorithmically constrained rules resists such consolidation. Immutable protocols eliminate the strategic option of political manipulation, compelling all actors to engage in productive economic calculation rather than lobbying or signalling. Drawing on Austrian capital theory and formal game-theoretic insights, this work demonstrates that rule stability is not a procedural convenience, but an existential prerequisite for decentralised coordination, long-horizon investment, and sustainable system governance.

\section{Time Preference and Institutional Frameworks}

Time preference lies at the core of intertemporal economic coordination, determining the willingness of individuals to defer present consumption for future returns. Yet this disposition is not purely psychological—it is shaped and constrained by the institutional framework within which economic actors operate. A credible, rule-bound environment encourages lower time preference by making long-term planning rational, while institutional volatility induces myopia and short-termism. In the context of Bitcoin, the protocol functions as a meta-institution: its perceived mutability or immutability profoundly influences the strategic horizons of miners, developers, and entrepreneurs. When rules are stable, actors can engage in roundabout, capital-intensive strategies that resemble Böhm-Bawerk’s temporal structure of production. When rules are subject to discretionary change, that same structure collapses into speculative churn. Thus, the time structure of economic planning in blockchain systems is not an exogenous given, but an emergent property of institutional credibility. This section integrates Austrian theory with dynamic game models to formalise how protocol stability anchors low time preference and rational economic calculation, establishing the preconditions for sustainable digital economies.

\subsection{Böhm-Bawerk and Temporal Capital Structure}

Eugen von Böhm-Bawerk’s theory of capital and interest rests upon the principle of “roundaboutness”—the idea that more capital-intensive and temporally extended production processes yield greater output but require lower time preference to be viable \cite{bohm1891capital}. Longer production periods are justified when individuals value future goods sufficiently to defer consumption. This temporal structure of capital accumulation is foundational to economic development: productive efficiency arises not from immediate gratification but from the disciplined deployment of resources over extended time horizons, enabled by institutional stability and consistent expectations.

This insight is profoundly relevant to Bitcoin mining, which constitutes a paradigmatic example of a roundabout production process in a digital context. Miners make substantial upfront capital investments in specialised hardware, access to stable energy supplies, and infrastructure for network connectivity. These sunk costs are undertaken with the expectation of future returns—specifically, the receipt of block rewards and transaction fees over a sustained operational horizon. The economic calculus guiding such decisions presumes that the protocol’s reward structures and validation rules will remain stable enough to render the investment rational. In this light, the miner mirrors Böhm-Bawerk’s capitalist-entrepreneur: one who allocates present goods to secure future outputs through a temporally extended and capital-intensive process.

However, if protocol rules are mutable—subject to discretionary forks, reward schedule alterations, or developer interventions—the foundation for such roundabout production collapses. A miner’s ability to calculate the marginal return on capital is impaired, truncating their strategic horizon. This transforms what should be a long-term productive investment into a speculative gamble. In Böhm-Bawerkian terms, institutional mutability raises effective time preference, reconfiguring the capital structure toward shorter, less efficient cycles. Thus, the analogy is not merely descriptive: Bitcoin mining under a credible, immutable protocol is the digital equivalent of Böhm-Bawerk’s temporal capital structure, while discretionary protocol governance injects distortions functionally equivalent to time-preference manipulation in fiat economies.

By framing miner behaviour within this Austrian temporal schema, we recast the design of blockchain protocols not as mere code optimisation but as institutional rule-setting that determines whether capital-intensive coordination is economically feasible. The critical question becomes not how to maximise hash rate per watt, but whether the rule environment supports the very formation of roundabout production structures.

\subsection{Misesian Praxeology and Economic Calculation}

Ludwig von Mises’s praxeology centres on the axiom that human action is purposeful behaviour aimed at the alleviation of felt uneasiness through the selection among scarce means. Central to this theory is the notion of economic calculation, which Mises defines as the ability of actors to allocate resources rationally based on the anticipated outcomes of their choices within a stable institutional framework \cite{mises1949human}. For Mises, calculability is not a mere convenience but a necessary condition for coordinated social planning, entrepreneurial judgement, and capital formation. Where rules are firm, individuals can act with confidence in the intertemporal structure of production; where rules are arbitrary or shifting, rational action dissolves into speculation.

Bitcoin, when conceived as a fixed-rule monetary and transaction validation protocol, functions analogously to a Misesian institutional framework: it facilitates calculation not by central planning but through the predictability of its algorithmic constraints. Miners allocate capital, estimate return horizons, and enter long-term commitments on the presumption that the rule set governing block validation, reward schedules, and network participation will remain unchanged. The credibility of these rules establishes the calculable environment within which praxeological coordination is feasible.

However, when protocol changes are introduced ex post—whether via contentious forks, discretionary soft changes, or non-transparent governance structures—this calculability is vitiated. Actors can no longer rely on present rules to hold in the future, and thus cannot perform genuine economic calculation. Instead, they are forced into second-order speculation about the intentions and influence of protocol developers or majority factions. This is not mere technical risk; it represents a collapse in the very logic of praxeological action. In Misesian terms, discretionary protocol mutability introduces institutional uncertainty, eliminating the possibility of rational economic planning and replacing entrepreneurial foresight with reactive opportunism.

Thus, rule stability is not a technical footnote but the fulcrum upon which economic rationality pivots. A protocol that shifts capriciously or under pressure ceases to be an institutional structure and becomes a game of discretionary power. The implications are profound: only under a credible commitment to rule permanence can the Bitcoin system function as a calculable order for entrepreneurial action. As Mises made clear, it is not the presence of computation or even prices that makes calculation possible, but the immutable framework of rules within which actors interpret those prices and deploy capital accordingly.

\subsection{Protocol Immutability as Institutional Anchoring}

Bitcoin’s original design embodies more than a technical architecture; it constitutes a synthetic constitution—an embedded institutional framework that governs the rules of economic engagement within the system. At its core, the protocol is not a fluid codebase subject to continual discretionary optimisation, but a foundational rule-set that defines the permissible bounds of action, expectation, and coordination among participants. Its stability serves the same anchoring function as a legal constitution: it allows actors to form long-term plans, to enter binding contracts in good faith, and to allocate capital based on reliable anticipations of systemic behaviour. In this sense, Bitcoin’s protocol must be understood not merely as software, but as “protocol-as-institution”—a game-defining structure that frames all subsequent strategic interaction.

Where traditional game theory assumes fixed payoffs and immutable rules, institutional game theory acknowledges that the meta-rules governing the game may themselves be contested or unstable. In the case of Bitcoin, protocol mutability constitutes a form of institutional endogeneity: rule changes, especially those enacted through developer discretion or non-transparent governance processes, do not simply adjust parameters within a known game; they redefine the game itself. A shift in block size, hash function, or reward schedule is not equivalent to a rule update in chess—it is the reconfiguration of what constitutes a legal move, who defines the board, and how payoffs are determined. Such interventions alter the strategic landscape retrospectively, eroding trust in the protocol’s constitutional character.

This mutability introduces regime uncertainty, a concept developed in the Austrian tradition to describe environments in which actors cannot confidently anticipate the future structure of rules. In such settings, coordination collapses and economic calculation becomes speculative. Just as discretionary monetary policy undermines the time structure of production by injecting artificial signals, so too does discretionary protocol revision disrupt the long-range planning of Bitcoin miners and businesses. Rational actors, unable to rely on the persistence of the institutional framework, resort to short-term, extractive strategies—accelerating entropy rather than fostering durable value.

Hence, the immutability of the Bitcoin protocol is not a dogma of ideological purity; it is an economic necessity grounded in institutional theory. It is the very condition that renders economic interaction calculable, cooperative behaviour rational, and roundabout capital allocation feasible. To forsake immutability is not merely to change the rules, but to destroy the game. In this light, protocol governance is not a matter of engineering optimisation, but of constitutional legitimacy.

\section{Strategic Behaviour of Miners under Rule Instability}

When institutional constraints governing economic interaction are subject to discretionary revision, strategic behaviour evolves not merely at the margin, but in kind. Miners, as rational actors embedded in a non-cooperative game of probabilistic competition, condition their strategies on the credibility of the underlying protocol. In stable systems, equilibrium dynamics reflect forward-looking optimisation—hash power is allocated to maximise discounted returns over a known horizon, often incorporating sunk capital costs and future fee projections. However, in unstable or mutable environments, the equilibrium contracts toward myopic extraction: not because miners cease to be rational, but because the game ceases to be formally stationary. Rule instability transforms the underlying payoff matrix, invalidating strategies predicated on continuity, and giving rise to behaviours that arbitrage uncertainty rather than invest in reliability. Such settings favour tactics that would be suboptimal in a credibly bound game—opportunistic forking, censorship-as-signal, and fee sniping become dominant not due to misaligned incentives, but because the structure of the game itself is corrupted. As the institutional scaffolding dissolves, the Nash equilibria recalibrate toward defection and resource dissipation, and the system’s ability to sustain intertemporal coordination deteriorates accordingly.

\subsection{Mining as Intertemporal Strategic Action}

Bitcoin mining is fundamentally an intertemporal act: it requires a present sacrifice of capital and energy in exchange for uncertain future returns. The miner commits resources—both tangible (ASIC hardware, electrical input) and intangible (network knowledge, technical maintenance)—to a process where rewards, in the form of block subsidies and transaction fees, are probabilistic and delayed. This structure mirrors the Austrian conception of investment as a temporally extended production process, whereby capital is allocated toward more “roundabout” methods in anticipation of a greater yield over time \cite{bohm1891capital}. Böhm-Bawerk’s insight that production becomes more productive through increased temporal structure finds a direct analogue in mining, where the decision to enter or exit, scale operations, or hedge exposure hinges on time preference and expectations of protocol consistency.

In Austrian theory, the viability of roundabout investment depends on the entrepreneur’s confidence in the institutional framework: whether the signals provided by relative prices and interest rates reflect genuine scarcity and temporal valuation. Mining, as entrepreneurial action, is no different. The profitability of a mining venture depends not only on current market conditions but on the assumed persistence of the rules—block sizes, difficulty adjustment, subsidy schedules—that shape the game’s payoff matrix. If these institutional constraints are mutable, or if developers and dominant factions can rewrite the protocol in response to exogenous pressures, then the rational horizon of mining contracts dramatically. What was an act of long-term planning becomes a gamble in a politically unstable environment. In such a context, even risk-neutral actors will exhibit behaviour characteristic of high time preference—not because they discount the future irrationally, but because the future is rendered institutionally indeterminate. Thus, mining must be understood as a praxeological category of investment shaped by institutional expectations, where intertemporal strategy and capital theory intersect most acutely.

\subsection{Short-Termism as Rational Adaptation}

In an institutional environment characterised by rule mutability, high time preference is not a behavioural anomaly but a rational adaptation. When the ground rules of the protocol are subject to discretionary revision, whether through opaque developer influence or ad hoc community consensus, the future becomes an unreliable domain for capital planning. For miners, whose operations hinge on precise anticipations of future reward schedules, transaction fee dynamics, and protocol parameters, such uncertainty compresses strategic horizons. As the reliability of long-term payoff structures erodes, rational agents default to immediate optimisation—extracting what can be gained now before the game itself changes.

This manifests empirically in phenomena such as empty block mining, where miners prioritise block propagation speed over fee accumulation; orphan fishing, where miners attempt to maximise returns by strategically exploiting propagation delays to override rival blocks; and transaction filtering, where potentially high-fee transactions are ignored in favour of censoring or prioritising structurally expedient ones. Each of these behaviours is economically perverse from a system-wide efficiency standpoint but individually rational under conditions of protocol-level indeterminacy. They represent tactical manoeuvres within a dynamic game whose parameters cannot be assumed stable—a situation akin to institutional regime uncertainty in Austrian political economy.

Thus, what may appear as irrational opportunism or moral failure in miner conduct is better understood as a form of constrained rationality. In a mutable protocol environment, long-term strategies become structurally penalised, while short-term extraction is rewarded not only economically but epistemically: it requires fewer assumptions about the continuity of institutional rules. Protocol mutability, therefore, does not merely distort incentives; it generates a selection pressure that systematically favours high time preference actors, accelerating the breakdown of trust and capital structure within the ecosystem.

\subsection{Rule Manipulation and Rent-Seeking}

In a system governed by immutable rules, agents optimise behaviour within clearly defined constraints. However, when protocol rules are mutable—subject to discretionary amendment by informal collectives or core developers—the economic structure ceases to resemble constrained optimisation and instead becomes a recursive contest over the rule-set itself. This engenders a second-order game wherein rational actors engage not only in mining, but in the strategic shaping of the rules that govern mining. The result is the emergence of rent-seeking behaviour, as miners and politically-connected stakeholders divert resources from productive investment toward institutional manipulation.

From the standpoint of Austrian economics, such behaviour corrodes the epistemic function of market prices and obliterates the possibility of genuine economic calculation. \citet{mises1949human} identified calculability as the cornerstone of rational planning, possible only within a stable institutional framework that permits entrepreneurial foresight. Rule mutability dissolves this framework, replacing calculable order with procedural indeterminacy. \citet{hayek1973law} further argued that spontaneous order depends on general, abstract rules which agents can internalise over time. When these rules become contingent upon political influence or shifting social norms, the resulting environment is not a market but a contest for control.

Game theory formalises this degeneration. In a fixed-rule environment, mining can be modelled as a repeated game with known payoffs, where equilibria correspond to cooperative or defecting strategies over long time horizons. When the rules themselves are variable, however, the game transforms into a “hypergame” \citep{bennett1977hypergames}—a game in which players are simultaneously playing and redefining the game. This recursive structure destabilises equilibria: players must not only consider current payoffs but the evolving meta-strategies of rule formation. Those best positioned to influence perception, consensus, and governance rituals can capture disproportionate benefits, irrespective of computational or economic efficiency.

Such dynamics have empirical analogues. Strategic advocacy for protocol changes that favour specific mining strategies (e.g., fee market design, transaction ordering mechanisms, or block size limitations) represents a shift toward rent-seeking. This is analogous to what \citet{buchanan1962calculus} termed “constitutional economics,” where political actors mould the rules for their own gain, distorting the collective rationality of the system. The blockchain becomes less a distributed ledger and more a discretionary polity—a system in which power is not constrained by constitutional anchoring but is instead wielded by coalitions engaged in narrative dominance.

The resulting institutional entropy erodes long-term investment. Rational miners internalise the heightened risk of rule-change and recalibrate their discounting behaviour accordingly. As \citet{lachmann1956capital} emphasised, capital formation is inseparable from institutional predictability. The more uncertain the future institutional regime, the higher the effective time preference, and the greater the incentive for short-run, extractive strategies. Mining strategies drift from optimisation of resource deployment to exploitation of political vectors. This reallocation of attention—from block production to block rule manipulation—marks the collapse of the entrepreneurial function in a spontaneous order.

In sum, rule mutability introduces a higher-order distortion: it incentivises meta-strategic behaviour incompatible with Austrian notions of decentralised coordination and game-theoretic stability. The system ceases to be a predictable institutional scaffold for entrepreneurial discovery and becomes a political economy of influence, narrative warfare, and structurally endogenous instability.

\section{Game-Theoretic Model of Miner Incentives}

Miner behaviour within blockchain systems can be formally captured through the lens of repeated games, wherein each round represents a block interval and players choose strategies conditional on both economic incentives and institutional parameters. Under a regime of protocol immutability, the strategic space stabilises: cooperation, honest propagation, and long-term investment become equilibrium behaviours sustained by credible deterrents to defection and the shared expectation of rule continuity. However, when protocol rules are mutable, the repeated game collapses into a hypergame—where players not only optimise within the existing payoff matrix but also compete to redefine it. This transformation introduces volatility into expected payoffs and undermines the coordination mechanisms essential for subgame-perfect equilibria. As the calculable future disintegrates, miners rationally discount cooperative strategies, favouring opportunistic extraction and meta-strategic positioning. The game evolves into a dual-layer contest: one over blocks, the other over the rules of the game itself.

\subsection{The Basic Miner Game}

Consider the mining landscape as a repeated game $\mathcal{G}$ where the set of players $N = \{1, 2, \dots, n\}$ represents miners, each endowed with computational power $\alpha_i$ such that $\sum_{i=1}^{n} \alpha_i = 1$. Each round (or stage game) corresponds to a block interval. The available strategies $S_i$ for each player include: (1) honest propagation (broadcasting found blocks immediately), (2) strategic withholding (e.g., selfish mining), and (3) collusive behaviour (coalitional deviation or cartel formation).

The payoff function $\pi_i : S \rightarrow \mathbb{R}$ depends on the probability of block success weighted by the player’s hashrate and adjusted by any additional gain or loss from deviation strategies. Crucially, each payoff is discounted over time by an intertemporal valuation parameter $\delta \in (0,1)$, reflecting the agent's time preference:
\[
U_i = \sum_{t=0}^{\infty} \delta^t \pi_i(s_i^t, s_{-i}^t \mid P_t)
\]
where $P_t$ denotes the protocol rule-set at time $t$. Under a regime of protocol immutability, $P_t = P_0$ for all $t$, and thus payoffs are stationary and expectations stable.

In this stable environment, miners can condition their strategies on the observed history of play. This permits the existence of subgame-perfect equilibria (SPE) that sustain cooperative norms. For instance, tit-for-tat or grim-trigger strategies may support honest propagation if the future value of cooperation outweighs the short-run gains from defection. As shown in the folk theorem for infinitely repeated games \citep{fudenberg1991game}, a wide set of cooperative outcomes can be sustained as equilibria when players are sufficiently patient ($\delta \to 1$).

Game-theoretically, immutability anchors the game’s equilibrium structure. The strategic space is bounded, the payoff matrix is stable, and expectations are aligned. This mirrors the Austrian requirement for calculability \citep{mises1949human}, where actors form plans based on reliable institutional scaffolding. Without rule drift, strategic deviations are costly in expected value terms due to the threat of future reversion strategies from other players.

Hence, the basic miner game under fixed protocol rules converges toward cooperative equilibria characterised by investment in hashpower, honest competition, and minimal meta-strategic interference. Rational miners internalise long-term benefits of order preservation and refrain from destructive short-term tactics. The Nash equilibria in this setting are not merely static, but intertemporally robust—sustained through credible threat of reversion and bounded rational expectation.

\subsection{Protocol Mutability as Game Redefinition}

Let the mining environment now be defined as a repeated stochastic game $\mathcal{G}^\prime$ where the rule-set $P_t$ evolves over time according to a probabilistic process. Define $\{P_t\}_{t=0}^{\infty}$ as a Markov chain over the space of protocol states $\mathcal{P}$, with transition probabilities $\mathbb{P}(P_{t+1} = p' \mid P_t = p) = \mu(p, p')$. Each realisation of $P_t$ reconfigures the payoff structure of the stage game. That is, for each $t$, the payoff function $\pi_i^t$ is contingent on the prevailing protocol rules:
\[
\pi_i^t = \pi_i(s_i^t, s_{-i}^t \mid P_t).
\]

In contrast to the fixed-rule setting, miners now face a stochastic environment where the payoff matrix is non-stationary. This transforms the game from a stable repeated interaction to a regime of shifting equilibria. The result is a degradation of strategic coherence: expectations over future payoffs acquire variance. Let $\sigma^2(\pi_i^t)$ denote the variance of expected revenue due to protocol volatility. Then, as the entropy of the rule distribution increases (i.e., $\mathcal{H}(\mu)$ rises), so does the second-order uncertainty embedded in each strategic choice.

This mutability imposes a real cost. In intertemporal terms, miners discount not only through time preference $\delta$, but also through epistemic risk $\eta$ associated with unknown future states. The adjusted utility becomes:
\[
U_i^\prime = \sum_{t=0}^{\infty} \delta^t \mathbb{E}_{P_t}[\pi_i^t] - \eta \cdot \sigma(\pi_i^t),
\]
where $\eta > 0$ encodes aversion to protocol uncertainty. Even when direct payoffs remain positive, elevated volatility erodes expected value through risk-adjusted utility.

From a game-theoretic standpoint, protocol mutability redefines the game itself. The miner's strategy space must expand to include not only in-game tactics but also metagame positioning—hedging against possible future rule-sets, investing in political influence, and dynamically reallocating hashpower. This destroys the possibility of subgame-perfect equilibria grounded in long-term cooperation: no credible threat can anchor behaviour when the payoff landscape itself is fluid and exogenous.

Austrian economic calculation collapses under such conditions. As Mises argued, calculability presupposes institutional constancy \citep{mises1949human}; if the metric of evaluation—the rule-set defining returns—mutates arbitrarily, rational planning becomes incoherent. In effect, mutability introduces a form of regime uncertainty akin to that discussed by Higgs in macroeconomic environments \citep{higgs1997regime}. Protocol instability mimics political discretion: economic actors respond by shortening time horizons, reallocating from investment to speculation, and privileging adaptability over optimisation.

Therefore, protocol mutability does not merely complicate strategy—it dismantles the foundational epistemic conditions of economic action. Mining becomes not a contest of productivity, but of anticipatory manoeuvring within a recursive system of rule uncertainty.

\subsection{Equilibrium Instability under Mutability}

In a mining environment governed by stable protocol rules, repeated-game equilibria often admit cooperative outcomes sustained by credible threat strategies, such as tit-for-tat or grim-trigger punishments. The Folk Theorem assures us that for sufficiently high discount factors and reliable institutional scaffolding, cooperative equilibria are subgame-perfect and self-enforcing. However, once protocol mutability is introduced, the incentive structure underlying cooperation deteriorates.

Formally, let us define the repeated game $\mathcal{G} = \{P, \pi, \delta\}$, where $P$ is the protocol, $\pi$ the payoff matrix, and $\delta$ the discount factor. With mutable protocols, the payoff matrix becomes $\pi_t$, a stochastic function of $P_t$. If $\mathbb{P}(P_{t+1} \neq P_t) = \epsilon > 0$, then players cannot form stable expectations about $\pi_{t+k}$ for $k > 0$. Under this condition, the expected future benefit of cooperation—dependent on protocol continuity—shrinks relative to immediate gains from unilateral defection. When:

\[
\delta \cdot \mathbb{E}[\pi^{coop}_{t+1}] < \pi^{defect}_t,
\]

then defection becomes rational in the current subgame. As the volatility of protocol rules increases, the expected value of future cooperation converges to zero, and the stage game reduces to a series of myopic optimisations.

This induces what we term *defection spirals*: one miner’s rational short-termism becomes a signal to others that long-horizon strategies are unstable. Strategic miners, anticipating future defection, preemptively defect. This positive feedback loop leads to generalised non-cooperation, collapsing the protocol’s economic function as a coordination mechanism. The equilibrium strategy profile shifts from cooperative Nash equilibria to dominant strategies centred on extraction, censorship, and strategic orphaning.

Simultaneously, a second-order strategy set emerges. Rational actors begin to allocate resources not only to mining per se, but to *meta-strategies*—actions aimed at shaping the future trajectory of the protocol. These include lobbying influential developers, engaging in social media persuasion campaigns, participating in consensus signalling rituals, and even coordinating politically to manipulate perception of miner majority.

The result is an endogenous bifurcation of strategy space: $\Sigma = \Sigma_{mine} \cup \Sigma_{meta}$. Profitability in $\Sigma_{meta}$ becomes commensurate with or superior to $\Sigma_{mine}$ as rule volatility increases. As Hayek warned, when rules become malleable to political will, actors cease to specialise in productive activity and begin to specialise in political action \citep{hayek1944serfdom}.

Consequently, equilibrium in mutable protocols is not merely unstable—it is systematically unattainable. The game becomes one of recursive meta-influence, where payoff surfaces evolve faster than players can adapt, and cooperation becomes a dominated strategy. In such an environment, economic rationality gives way to performative signalling, and the institutional core of Bitcoin mining as a calculable order is eroded.

\section{Case Study: Bitcoin vs BTC Core}

The divergence between the original Bitcoin protocol and the BTC Core implementation provides an empirical instantiation of the theoretical dynamics discussed. Initially, Bitcoin operated under a fixed-rule framework: a stable protocol with known constraints, a defined block size, and predictable economic incentives. Miners, developers, and businesses could form expectations about future network behaviour and invest accordingly. However, with the rise of BTC Core governance, rule mutability became institutionalised. Protocol changes—such as the imposition of a 1MB block size limit, the introduction of replace-by-fee (RBF), modifications to mempool policy, and social enforcement of soft forks—altered not only the technical structure but also the strategic terrain. The result has been a reallocation of economic activity: mining power shifted to short-term opportunistic strategies, fee variance increased, and transaction capacity became externally constrained, encouraging off-chain substitutes like Lightning and custodial batching.

From a game-theoretic perspective, the BTC Core model redefines the miner’s game from one of long-term participation to one of short-term positional advantage. Influence over rule-making became a strategic asset, fostering political competition rather than entrepreneurial productivity. In Austrian terms, the calculability of the market order was compromised. Rational actors responded by shortening investment horizons, minimising sunk costs, and seeking rent not from validation but from meta-protocol leverage. By contrast, the original Bitcoin protocol—when enforced as immutable—allowed for long-range planning, large-scale infrastructure development, and a capital structure aligned with deferred reward. The divergence is not merely technical; it reveals the existential economic consequence of institutional design. Where rules are mutable, strategy decays into theatre. Where rules are fixed, strategy evolves into investment.

\subsection{Historical Divergence in Incentive Structures}

The divergence between the original Bitcoin protocol and the contemporary BTC Core implementation illustrates a paradigmatic shift in miner incentive structures, grounded not merely in technical modifications but in institutional realignment. Originally, Bitcoin’s design encouraged fee-based scaling through unrestricted block sizes and competition among miners to include as many high-fee transactions as possible. The game theoretic equilibrium incentivised maximising utility for users, aligning miner revenue with throughput expansion and long-term network utility.

Under the BTC Core regime, however, key protocol changes have altered these incentive structures. The imposition of an artificial block size limit—initially 1 MB and later formalised through soft forks such as Segregated Witness—has introduced an exogenous constraint on block composition. This has decoupled miner incentives from transaction volume, redirecting their focus from throughput optimisation to fee bidding games within a congested mempool \citep{bowe2017segwit}.

Further, the adoption of restrictive mempool policies (e.g. Replace-by-Fee, minimum relay fees) and increasingly discretionary relay rules have transformed the transaction propagation process into a controlled bottleneck \citep{corallo2016rbf}. Rather than functioning as a neutral network layer, mempool policy under BTC Core acts as a regulatory filter—privileging certain transaction profiles while excluding others based on dynamic fee thresholds or script composition. This introduces institutional bias into what was originally designed as a permissionless, content-neutral system.

Empirically, these changes have been accompanied by strategic behaviour incompatible with the assumptions of the original protocol. The prevalence of empty block mining during periods of high mempool congestion—despite the availability of lucrative fees—demonstrates a rational adaptation to altered incentives. In particular, miners respond not to user demand, but to the increased complexity and cost of transaction validation under restrictive rulesets. As block propagation delays become more costly, the Nash equilibrium shifts towards minimal validation strategies, undermining Bitcoin’s role as a transaction processing system \citep{carlsten2016fee}.

Institutionally, this reflects a drift from protocol-as-rule to protocol-as-policy. The mutable intervention of core developers—acting as discretionary governors of technical parameters—transforms Bitcoin’s coordination layer from an impersonal system of incentives into a politicised site of control. In Austrian terms, the calculability of future returns collapses; economic actors cannot plan capital deployment when the rule structure is subject to ongoing revision. Miner time horizons shorten, and entrepreneurial activity shifts from building scalable infrastructure to gaming the evolving ruleset.

This historical divergence is not merely technical. It represents a collapse in institutional coherence, where the originally envisioned spontaneous order has been supplanted by a cartelised governance model. The game miners now play is not the one encoded in 2009, but a mutable construct whose payoff matrix is determined by off-chain coordination among a narrow group of developers and signalling entities. In such a regime, incentive structures become opaque, and economic calculation is replaced by political speculation.

\subsection{Observable Miner Responses}

The behavioural response of miners to protocol mutability is not merely theoretical—it is empirically evident in measurable changes to mining patterns, fee markets, and infrastructure development. These observable outcomes correlate closely with interventions in the BTC Core rule framework and signal a rational adaptation to altered incentive landscapes.

First, shorter average block intervals during periods of fee pressure, despite the 10-minute target, suggest aggressive variance-minimising strategies. Mining pools routinely calibrate hashpower allocation to front-run competitors in anticipation of high-fee inclusion, often resulting in clusters of rapid block discovery followed by lulls. This reflects a shift away from smooth intertemporal optimisation towards burst-based extraction, a hallmark of rising time preference induced by rule instability. The predictability of revenue streams diminishes, and miners revert to myopic allocation strategies driven by near-term mempool states rather than long-run fee accumulation.

Second, increasing variance in fee revenue among blocks mined within short windows indicates structural disequilibrium. Under stable protocol conditions, competition should yield convergence towards marginal fee optimisation. However, when relay rules, transaction admission policies, and block size ceilings are mutable or ambiguous, the market for transaction inclusion becomes balkanised. Some miners opportunistically favour small, easily propagated blocks to minimise orphan risk, while others gamble on high-fee density. The result is a dispersion of payoffs reflecting not skill in block construction but differential exposure to exogenous protocol governance and selective relaying behaviour \citep{carlsten2016fee}.

Third, the growth of off-chain transaction substitutes—ranging from custodial wallets to Layer-2 constructs—illustrates market circumvention of unreliable base-layer guarantees. Users and businesses, unable to reliably estimate settlement times or fees under shifting policy, opt for trusted intermediaries. This reintroduces counterparty risk and collapses the decentralisation dividend originally promised by the protocol. From the miner’s perspective, this shrinkage of on-chain economic activity further compresses long-term revenue expectations, reinforcing the short-termism cycle. When protocol design fails to anchor calculable expectations, the ecosystem as a whole fragments into trust-minimised niches dominated by opaque coordination and platform risk.

Collectively, these observable miner behaviours constitute a diagnostic signal: they reflect not individual irrationality, but systemic adaptation to unstable institutional signals. In Austrian terms, the erosion of calculability transforms entrepreneurial mining from capital-guided discovery to risk-hedged speculation. In game theoretic language, miners cease to play the original repeated game and instead operate within a dynamic environment of shifting payoffs and emergent meta-strategies, leading to degraded cooperation and the rise of opportunistic equilibria.

\subsection{Austrian Diagnosis}

The empirical responses of miners to protocol mutability—strategic variance in block production, fee dispersion, and migration to off-chain channels—are best understood through the lens of Austrian capital theory and monetary economics. At their core, these behaviours reflect heightened discounting of future income streams and a systemic breakdown in the calculability required for rational economic planning.

In Austrian theory, the entrepreneur operates under uncertainty, but not chaos. Calculability—the ability to form coherent expectations about future conditions—is the foundation of economic coordination in a market order. As Mises emphasises, economic action presupposes a framework of stable institutions within which actors can interpret prices, forecast returns, and allocate capital accordingly \citep{mises1949human}. When that framework is disrupted—when rules are mutable and future returns are contingent on discretionary intervention—rational actors will necessarily discount the future more heavily. The time horizon of investment contracts. Long-term strategies give way to short-term speculation, as seen in the miners’ pivot from cumulative fee optimisation to opportunistic block construction.

This dynamic closely parallels the Austrian diagnosis of monetary instability in fiat regimes. When a central bank engages in discretionary monetary policy—altering interest rates, expanding base money, or signalling future interventions—the informational role of money is compromised. Economic actors, uncertain about the real value of future cash flows, reallocate resources towards consumption, arbitrage, or speculative hedging. The temporal structure of production is deformed: lower-order goods are favoured, and capital-intensive, roundabout processes are abandoned \citep{rothbard2009mystery}. Similarly, in a mutable blockchain protocol, the informational clarity of the ruleset is degraded. Miners, uncertain about future conditions of block size, transaction relay, or fee structures, abandon intertemporal optimisation in favour of tactics that maximise immediate extractable value.

The analogy is precise: just as discretionary monetary regimes induce higher time preference and misallocation in the broader economy, discretionary protocol governance induces high-frequency, short-horizon strategies among miners. What appears as irrational volatility is in fact a rational response to institutional erosion. The Austrian insight here is critical—time preference is not merely a psychological parameter, but a systemic response to the stability (or instability) of the institutional environment. Where calculability fails, the market process cannot function as a discovery procedure. In such an environment, entrepreneurial function devolves into political navigation, and capital becomes trapped in games of institutional arbitrage rather than productive coordination.

\section{Institutional Certainty and the Restoration of Rational Investment}

Rational economic action requires a horizon of expectation grounded in institutional stability. Without calculable frameworks, the entrepreneur cannot allocate capital across time, and the investor cannot assess risk beyond the present moment. In blockchain ecosystems, protocol immutability functions analogously to constitutional law—limiting the domain of permissible action and enabling the formation of stable expectations. This is not merely a technical constraint but the sine qua non of rational planning. Austrian capital theory teaches that roundabout production processes—those requiring deep time, deferred reward, and high capital commitment—emerge only where institutional trust exists. When protocol rules are subject to revision through informal consensus or discretionary override, the time structure of production collapses. Investment recedes, extraction accelerates, and the entire economic architecture devolves into immediacy and arbitrage. The restoration of rational investment within blockchain systems therefore hinges not on improved incentives or market signals alone, but on the reconstitution of rule certainty. Only under fixed institutional constraints can entrepreneurial foresight operate, transforming blockchain from an opportunistic game into a productive economic order.

\subsection{Reinstating Protocol as Fixed Law}

The restoration of protocol immutability is not merely a technical correction but a categorical economic necessity. In an environment where future expectations must underpin costly, long-term capital commitments—such as those undertaken by industrial miners—any deviation from predictable, rule-bound structure leads inexorably to systemic fragility. Reinstating the protocol as a fixed, non-discretionary foundation is essential to reconstitute the conditions of calculative action and to re-anchor entrepreneurial confidence. This is not a plea for rigidity for its own sake, but a recognition that in economic systems, certain constraints are constitutive, not optional.

Drawing from Menger’s foundational insight, law is not a product of deliberate invention but the gradual crystallisation of practices that emerge through voluntary exchange and institutional evolution \citep{menger1883investigations}. Protocol, in this sense, ought to function analogously to law: not a mutable set of preferences imposed by developers, but a stable framework emerging from, and reinforced by, the actions of market participants. When rules are subject to change through political rituals or discretionary override, they cease to serve their role as epistemic anchors. Entrepreneurs, particularly miners, must then shift resources not toward productive coordination but toward anticipating and manipulating future rule changes. The result is a degeneration of the system’s informational coherence—an erosion of the very mechanism that allows decentralised actors to converge upon shared expectations.

To treat the protocol as mutable is to misinterpret its function. It is not a parameter set to be optimised; it is the game’s constitution. By reinstating a fixed protocol, the system can again support long-range economic calculation, lower time preference, and high-capital investment. Stability in the rule framework enables the spontaneous order described by Hayek and Menger to flourish—one in which cooperation does not require direct communication, but unfolds through the mediated signals of a predictable institutional structure \citep{hayek1945use}. Without such fixity, strategic coherence collapses, and the system devolves into institutional warfare.

Thus, protocol immutability is the precondition for any credible game-theoretic equilibrium. It is not the preservation of past code per se that matters, but the reconstitution of protocol as law—emergent, general, and fixed in its capacity to coordinate expectations. This transforms mining from a tactical arbitrage game into a strategic, time-anchored pursuit of value. Without it, no meaningful market process can be said to exist.

\subsection{Re-aligning Game Equilibria}

When protocol rules are rendered stable and immune from discretionary revision, the strategic environment for miners undergoes a fundamental transformation. The game ceases to be a recursive negotiation over institutional form and returns to its original structure: a repeated game of production under known constraints. Under these conditions, the equilibrium set stabilises. Long-term cooperative strategies, such as honest block propagation and investment in network infrastructure, once again become rational. Agents need no longer hedge against institutional caprice, nor divert resources into rent-seeking meta-strategies; instead, they optimise against a shared, immutable rule set that renders expectations calculable.

From a game-theoretic perspective, fixed protocols convert a hypergame—where players simultaneously play the game and alter its rules—into a standard repeated game with well-defined payoffs. In such a game, tit-for-tat and grim-trigger strategies regain their efficacy as deterrents against deviation \citep{fudenberg1991game}. Trust, though never explicit, becomes embedded in the institutional scaffold: rational agents assume that their counterparts face the same incentive landscape and can thus coordinate without direct collusion. Cooperation ceases to be a fragile artefact of overlapping interests and becomes a dominant strategy in the presence of credible commitments.

This has profound implications for governance. Protocol-level stability limits the political surface area of the system. It renders consensus governance not a mechanism for change, but a form of constitutional defence. Just as the rule of law in liberal democracies does not imply constant revision of legal codes, but their procedural stability, so too must protocol governance reorient toward restraint, not activism. The Austrian emphasis on institutional calculability is instructive here: economic actors must operate under rules that are not only known, but expected to remain in force over their planning horizons \citep{rothbard2009man}.

In practical terms, this re-alignment fosters a mining ecosystem oriented toward capital formation, long-term planning, and entrepreneurial experimentation. The block reward, when situated within a stable institutional frame, functions not as a bounty subject to political renegotiation, but as a reliable vector for investment return. Time preference declines, cooperative equilibria emerge, and the system regains its economic coherence. By reinstating fixed rules, we do not merely preserve the past—we construct the only viable game that supports the emergence of spontaneous order.

\section{Institutional Certainty and the Restoration of Rational Investment}

\subsection{Endogenous Time Preference under Institutional Noise}

In classical Austrian economics, time preference is treated as a subjective, ordinal ranking of present versus future satisfactions. Mises viewed this preference as a universal category of action, but also one that expresses itself variably depending on institutional and informational conditions. Building on this, we introduce the concept of endogenous time preference adjustment: a dynamic reaction of temporal valuation not to internal psychological shifts, but to external volatility in institutional frameworks. In the case of blockchain miners, the institutional framework is the protocol—its rules, governance procedures, and perceived stability. When those rules become mutable, discretionary, or unpredictably modifiable, agents adjust their subjective time preference upwards as a rational response to increasing uncertainty.

Mathematically, we model this effect by allowing the discount rate \( \rho \) in a standard intertemporal utility function to become a function of institutional noise \( \eta(t) \), where \( \eta(t) \) captures the perceived variance in protocol stability at time \( t \). Thus, the effective discount function becomes:

\[
\delta(t) = e^{-\int_0^t (\rho + \eta(s)) ds}
\]

Here, \( \eta(s) \) represents a stochastic process reflecting rule-change volatility, such that increases in protocol-level uncertainty directly raise the effective discount rate over time. In a regime of stable, immutable rules, \( \eta(s) = 0 \), and the agent discounts at their baseline subjective rate \( \rho \). But as rule instability increases—e.g., through unexpected hard forks, mempool policy shifts, or sudden developer interventions—\( \eta(s) > 0 \), amplifying the discounting effect and thereby raising the agent's preference for immediate over deferred rewards.

This transforms the economic environment from a domain of rational intertemporal coordination into one of high-frequency extraction. Capital-intensive behaviour, such as purchasing and deploying ASIC hardware with multi-year amortisation schedules, becomes irrational under elevated \( \eta(t) \). The calculative horizon shortens not because of irrationality, but because the future becomes structurally illegible. As institutional reliability decays, so too does the viability of roundabout production processes. The implications extend beyond miner conduct to entrepreneurial activity writ large: all planning that depends on rule-based extrapolation collapses into speculation once protocol noise exceeds calculability thresholds.

In Austrian terms, the institutional environment ceases to function as a coordination-enabling framework and becomes instead a volatility-inducing parameter. The entrepreneur no longer anticipates future states through deductive reasoning from known rules, but must hedge against politically-induced protocol drift. This shift from calculative action to speculative adaptation undermines the entire epistemic function of money and contractual engagement. Where Hayekian order relies on the feedback structure of stable institutions and dispersed knowledge, institutional noise generates a regime where expectations become decoupled from productive foresight, and economic action collapses into a reactive posture of defensive liquidity.

\subsection{Strategic Investment in High-Variance Games}

In economic environments characterised by institutional volatility, strategic investment transforms from a game of productive deployment into a high-variance contest of timing, signalling, and rent-seeking. Traditional capital theory, particularly within the Austrian framework, emphasises roundaboutness: the extension of production processes across time in pursuit of greater output through delayed gratification. This model presupposes calculable institutional continuity. Yet when the foundational rules of engagement—such as those embedded in blockchain protocols—are mutable, the capital investment landscape becomes distorted. In game-theoretic terms, the payoff matrix of mining is no longer defined ex ante. Instead, the game becomes reflexive: each move by a miner is influenced not only by the current parameters but by the anticipated manipulation or redefinition of those parameters by other strategic actors.

Formally, consider a repeated game \( G_t \) where the transition probabilities between stages and the utility outcomes at each node are functions of \( \theta_t \), a stochastic variable representing protocol state. In stable systems, \( \theta_t = \theta \) for all \( t \), rendering the game stationary. Under mutability, however, \( \theta_t \sim \mathcal{N}(\mu, \sigma^2) \), where \( \sigma^2 \) measures protocol volatility. This non-stationarity invalidates standard backward induction and disrupts subgame-perfect equilibrium paths. Instead of seeking long-run utility maximisation, rational actors begin to privilege strategy profiles that hedge against protocol drift—strategies that minimise variance rather than maximise mean payoff.

This results in a preference for liquidity over capital lock-in, a retreat from deferred reward structures, and an increase in defection-type strategies. Mining behaviours adapt accordingly: hash withholding, zero-confirmation exploitation, and collusion over block order are not necessarily malicious deviations from protocol norms, but rational responses to strategic uncertainty. Moreover, the incentives to engage in meta-strategic activity—such as signalling allegiance to influential developer factions or participating in social consensus manipulation—become endogenous to the game itself. In such environments, strategic positioning eclipses productive capital allocation.

Austrian critiques of central planning have long warned of the calculational chaos introduced when prices or rules are centrally adjusted rather than emerging organically from market interaction. Here, the analogy is exact. When protocol rules are subject to ad hoc change by a privileged priesthood—whether they be foundation developers or social consensus gatekeepers—the information signals upon which rational investment depends are corrupted. The game’s internal logic becomes inconsistent across iterations, and the investor is reduced to a speculator in governance outcomes.

Thus, the emergence of strategic investment in high-variance games reveals a broader institutional pathology: that of the discretionary order. In such a system, the return on capital is decoupled from productive contribution and reattached to social alignment with power centres. As equilibrium investment migrates away from entrepreneurship and towards political rent-seeking, the system’s economic capital base erodes, and the very sustainability of the network is placed in jeopardy. Only through rule fixity can the variance be collapsed, calculability restored, and investment rechannelled into cooperative equilibria anchored in long-horizon planning.

\subsection{Capital Roundaboutness and Deferred Payoff Structures}

Böhm-Bawerk’s conception of capital roundaboutness articulates a profound insight into the temporal structure of production: namely, that longer, more capital-intensive production processes—though delayed in fruition—yield proportionally higher returns due to their layered integration of intermediate goods and temporal leverage. This insight, central to Austrian capital theory, finds a structural analogue in Bitcoin mining, wherein actors commit significant upfront capital—ASIC hardware, infrastructure, energy contracts—with the expectation of distributed future payoffs in the form of block rewards and transaction fees. These deferred payoff structures are not anomalies but exemplars of economic roundaboutness: they embody intertemporal coordination within a rule-bound environment.

However, roundabout investment strategies presuppose a predictable institutional framework. The miner’s calculus—like the capitalist’s—is contingent upon the constancy of rules across time. When protocol mutability is introduced, the temporal chain that justifies roundaboutness fractures. Deferred payoff structures become discounted not merely by subjective time preference \( \rho \), but by institutional uncertainty \( \eta(t) \), as discussed previously. The effective cost of capital thus rises not from real interest rates, but from political entropy embedded in the system’s protocol governance. Rational actors, facing this composite discounting, shift investment strategies toward immediacy and liquidity, forsaking roundabout commitments for extractive proximity.

We can formalise this transformation. Let the expected net present value (NPV) of a mining investment \( I \) be given by:

\[
\text{NPV}(I) = \sum_{t=1}^{T} \frac{E[R_t]}{(1 + \rho + \eta(t))^t} - C_0
\]

where \( E[R_t] \) denotes expected return in period \( t \), \( \rho \) is subjective time preference, \( \eta(t) \) reflects institutional noise, and \( C_0 \) is the upfront capital outlay. As \( \eta(t) \) increases, the present value of long-term payoffs collapses, rendering roundabout strategies uneconomic. The investment frontier contracts inward, privileging short-term strategies that minimise exposure to future institutional risk.

This dynamic aligns precisely with Menger’s insight that the formation of institutions—especially legal and monetary—is an organic, emergent process, not one subject to centralised authorship or ad hoc revision. Deferred payoff structures in blockchain systems can only function when the institutional environment mirrors this emergent constancy. When protocol development becomes politicised, discretionary, or opaque, the feedback loop that sustains long-horizon entrepreneurial action is broken.

Empirically, this manifests in the compression of miner investment cycles. Whereas early Bitcoin miners operated under the assumption of multi-year recoverability anchored in protocol fixity, modern miners in mutable regimes exhibit behaviours characteristic of high-volatility financial arbitrage: rapid hardware turnover, minimised sunk costs, and strategic responsiveness to short-term shifts in policy. The long arc of roundaboutness is thus truncated, and with it, the capacity of the system to support meaningful capital formation. Only by restoring protocol immutability can deferred payoff structures regain calculative viability and re-align miner incentives with the long-term sustainability of the network.
\subsection{Commitment Devices and Institutional Anchoring}

In environments of radical uncertainty, rational agents seek anchoring mechanisms that reduce the volatility of strategic expectations. Commitment devices serve this anchoring role by constraining future actions, thereby reducing strategic ambiguity and enabling the emergence of equilibrium behaviour across time. Within blockchain systems, the fixedness of the protocol functions as such a commitment device: it assures miners that today’s investment assumptions will not be invalidated by tomorrow’s discretionary changes. This assurance forms the foundation for entrepreneurial calculation, allowing participants to formulate coherent expectations about future payoffs and plan accordingly.

Drawing on Schelling’s foundational insights into strategic commitments, the power of a commitment device lies precisely in its irreversibility. A protocol that is resistant to arbitrary revision mimics the function of a constitution in the political realm: it constrains future discretion and provides a credible signal to all participants that strategic moves made today will not be undercut by exogenous institutional volatility. This is not merely a governance feature but an economic necessity. Without such anchoring, time preference endogenously rises, as agents rationally discount future payoffs that are perceived as uncertain, malleable, or politically contingent.

In the language of repeated games, fixed protocols transform the underlying game from a stochastic hypergame into a Markov stationary process with clear state transitions. This transformation allows for the emergence of subgame-perfect equilibria sustained by cooperative strategies, such as honest mining and network reinforcement. Conversely, when commitment devices are absent or unreliable—as in systems governed by discretionary development teams or mutable consensus rituals—the game devolves into a regime of recursive renegotiation. Agents can no longer optimise over time; they must hedge against institutional betrayal.

The Austrian school’s emphasis on institutional predictability—seen in Mises’s concept of economic calculation and Hayek’s theory of spontaneous order—makes clear that calculability is not a luxury but a precondition for rational economic behaviour. Institutions that serve as commitment devices, by remaining fixed and impersonal, provide the framework within which entrepreneurial judgement operates. The moment protocol rules become subject to discretionary amendment, they cease to function as anchors and instead become strategic variables within a higher-order game of influence and signalling.

Blockchain systems that lack robust commitment mechanisms inevitably trend toward governance centralisation. The discretion once reserved for external regulators becomes internalised in the developer class, whose capacity to mutate protocol parameters introduces uncertainty not unlike that of fiat regimes. This erosion of rule-predictability forces miners to either become political actors themselves or surrender long-horizon strategy in favour of immediate, opportunistic engagement. Thus, commitment devices are not ideological artefacts but structural necessities. They tether strategic time horizons to a stable institutional frame, re-enable deferred investment, and restore the foundational calculability required for sustainable network growth.

\subsection{A Meta-Model of Protocol Rule-Sets as Economic Institutions}

To conceptualise blockchain protocols as economic institutions, one must shift from viewing them as mere technical artefacts to recognising them as embedded rule-sets structuring strategic behaviour. In this framing, a protocol is not simply code; it is a meta-institution that defines permissible action, mediates conflict, and enables the calculative coherence necessary for complex coordination. From an Austrian perspective, institutions are emergent constraints that facilitate the formation of expectations and the realisation of entrepreneurial plans. When protocols are treated as mutable engineering projects, rather than institutional commitments, their economic role is fundamentally compromised.

Let us formalise this with a meta-model. Consider a protocol \( P \) as a set of rules \( \{r_1, r_2, \dots, r_n\} \), each of which shapes the payoff space \( \Pi \) of the economic game \( G \) played by miners. If the rule-set \( P \) is stable, then the payoff matrix \( \Pi_P \) remains invariant across periods \( t \in [1, T] \), permitting the derivation of equilibrium strategies under repeated game conditions. However, if \( P \) is itself a stochastic variable—subject to amendment with some probability \( \delta > 0 \)—then \( \Pi_P(t) \) becomes time-dependent and path-contingent. In such a system, players not only select strategies within \( G \) but must also engage in meta-strategic behaviour aimed at influencing \( P \) itself.

We define this as a two-tiered game:

\begin{enumerate}
  \item The base game \( G_P \): strategic interaction under a given protocol \( P \), with known payoff structure \( \Pi_P \).
  \item The meta-game \( M \): a contest over the evolution of \( P \), wherein agents allocate capital, narrative effort, and coordination to shape or forestall rule changes.
\end{enumerate}

In the presence of high \( \delta \), the meta-game \( M \) dominates. Rational agents reallocate capital away from productive strategies in \( G_P \) toward rent-seeking strategies in \( M \). This introduces institutional endogenous risk, which cannot be hedged through ordinary market means. The entire protocol structure becomes less akin to a constitutional order and more akin to a parliamentary system with no fixed-term limits or separation of powers.

From this vantage, we can model institutional integrity as a function \( \Theta(P) \) that captures the protocol's resistance to discretionary change. A high \( \Theta \) correlates with calculability, low time preference, and long-horizon investment. A low \( \Theta \) implies discretionary capture, recursive rent-seeking, and hyperbolic discounting. Hence, the protocol’s economic identity lies not in its codebase but in its rule permanence. When protocol governance lacks commitment mechanisms—whether cryptographic, legal, or socio-economic—the system devolves from a constitutionally ordered game to a recursive contest over rule authorship.

This meta-institutional lens reframes debates about block size, fee structure, or consensus mechanism as not merely technical or empirical concerns but as constitutional choices with deep economic consequences. Each change to \( P \) is a legislative act, and each act either strengthens or erodes the institutional coherence that undergirds calculability. Consequently, protocol design must prioritise institutional durability over technocratic responsiveness, recognising that the cost of mutability is borne not in immediate efficiency losses but in the silent erosion of strategic time.

\subsection{Implications for Broader Institutional Cryptoeconomics}

The insights derived from modelling blockchain protocols as economic institutions extend beyond Bitcoin or its derivatives and strike at the core of cryptoeconomic design. At stake is the broader theoretical architecture underpinning how decentralised systems mediate commitment, calculability, and trust over time. In environments where consensus rules are discretionary or readily modifiable—whether through informal developer coalitions, token-holder voting, or governance tokens—the very notion of credible commitment collapses. These systems increasingly resemble fiat economies with endogenous institutional noise, undermining the purported stability such technologies aim to provide.

In Austrian terms, the emergence of calculable order is not an automatic by-product of decentralisation; it requires institutional constraints that are both known and non-discretionary. Menger's conception of money, Mises's focus on economic calculation, and Hayek’s theory of spontaneous order all converge on the necessity of predictable institutional scaffolding. When blockchains mimic political systems—subject to majoritarian rule, ideological flux, or elite capture—their monetary and economic functions begin to mirror those of unstable states: subject to high temporal discounting, opportunism, and strategic redefinition.

This has profound consequences for cryptoeconomic systems that depend on long-term participation: staking, stablecoins, decentralised financial instruments, or smart contract execution platforms. Each of these layers presupposes the calculability of the underlying institutional environment. If base-layer rules are mutable, then the entire stack inherits instability. Investors cannot model expected returns; application developers cannot price future states; users cannot form stable expectations about security guarantees. In effect, the system loses its ability to function as a medium for long-term coordination.

Game theoretically, these systems transition from games of incomplete information to games of indeterminate structure. In such a setting, the design of mechanisms—be it fee markets, incentive pools, or penalty structures—fails to guarantee equilibrium behaviour because the game’s boundary conditions remain contestable. This creates a reflexive problem: the very actors participating in the game are simultaneously incentivised to redefine the game’s structure. The Nash equilibria of such systems are path-dependent and politically fragile.

To restore institutional coherence, cryptoeconomic design must abandon the fetish of governance dynamism in favour of a jurisprudential model of protocol law. This means elevating rules to the status of commitments rather than preferences. Cryptographic mechanisms must be complemented by normative constraints—legal, social, and structural—that lock in institutional expectations over time. The protocols that survive will not be the most flexible or the most participatory, but those that constrain arbitrary change, restore the calculability of economic interaction, and reduce the institutional entropy that currently defines much of the blockchain landscape.

Thus, institutional cryptoeconomics must be re-centred on the logic of rule-bound calculability. Rather than framing consensus protocols as upgradable software layers, they must be understood as constitutions: meta-institutions whose legitimacy and utility derive from their fixedness. This is not a technocratic proposal but a foundational one. Without fixed rules, no coherent theory of value, investment, or coordination is possible in the cryptoeconomic domain.

\section{Conclusion}

This paper has developed a hybrid theoretical synthesis integrating Austrian capital theory with formal game-theoretic analysis to evaluate the strategic consequences of protocol mutability in blockchain systems. From the Austrian perspective, we have shown that mutable rules elevate time preference, thereby undermining the long-range calculability necessary for entrepreneurial coordination and capital investment. The epistemic conditions required for rational planning—clear expectations, institutional stability, and consistent norms—are obliterated when protocol rules are subject to discretionary alteration. This mirrors the destabilising effects of monetary unpredictability under fiat regimes, where economic calculation is distorted by political tampering.

From a game-theoretic lens, mutable protocols dissolve cooperative equilibria. The mining game becomes unstable, as rational actors abandon long-term strategies in favour of short-term extraction and meta-strategic manipulation. Equilibrium degeneracy, defection spirals, and institutional gaming emerge as natural consequences of uncertainty in payoff structure. The system transitions from a repeated game of productive action to a hypergame of discretionary influence.

The solution is not to redesign incentive mechanisms ad infinitum but to reinstate protocol immutability as the constitutional core of the system. A fixed rule-set restores the calculability required for low time preference, aligns incentives around productive behaviour, and re-establishes the cooperative equilibria that make decentralised verification viable. In this sense, protocol immutability is not merely a technical feature—it is the precondition for any enduring economic order within blockchain-based systems. Without it, entrepreneurship withers, rent-seeking dominates, and the promise of distributed consensus collapses into procedural theatre.

\subsection{Fixed Rules as the Foundation of Cryptoeconomic Rationality}

The foregoing analysis reaffirms that the calculability necessary for rational economic coordination—central to both Austrian theory and game-theoretic models—depends not on decentralisation per se, but on the permanence and inviolability of institutional rules. Protocol mutability dissolves the temporal anchor of entrepreneurial planning, replacing investment with speculation and cooperation with political manoeuvring. When the structure of the game is itself contestable, defection becomes rational, discounting accelerates, and capital migrates toward rent-seeking rather than productive commitment.

This pathology is not confined to Bitcoin or its contested variants. It manifests wherever cryptoeconomic systems treat rule evolution as a dynamic feature rather than a foundational liability. The broader cryptoeconomic space now resembles a landscape of institutional instability—where consensus is performative, governance is discretionary, and incentives are unstable. If blockchain systems are to fulfil their promise as coordination infrastructures for the digital economy, they must return to first principles: fixed rules, non-discretionary enforcement, and institutional coherence.

To build enduring systems, cryptoeconomic architects must reject the false dichotomy between flexibility and ossification. Instead, they must embrace a constitutional model of protocol law, wherein the base rules are insulated from manipulation and interpreted as commitments, not preferences. Only in such a setting can the miner behave as an entrepreneur, not a political agent; only then can capital structure reflect roundabout investment and long time horizons; and only then can decentralised systems provide the stability upon which calculative rationality and spontaneous order depend. Fixed rules are not a constraint on innovation—they are the precondition of it.

\bibliographystyle{apalike}
\bibliography{references11}

\end{document}